# Photovoltaic Performance of a Rotationally Faulted Multilayer Graphene/n-Si Schottky Junction

Hojun Im[*] and Masahiro Teraoka

*Graduate School of Science and Technology, Hirosaki University, Hirosaki 036-8561, Japan*

**ABSTRACT**

We report the fabrication and photovoltaic performance of a rotationally faulted multilayer graphene (rf-MLG)/n-Si Schottky junction device. A thickness-controlled rf-MLG is synthesized using a 5 μm Ni foil catalyst via the chemical vapor deposition method and transferred to the n-Si substrate via a polymer-free process, enabling facile and cost-effective fabrication. The device demonstrates an ideality factor of 1.67, a rectification factor of approximately $4 \times 10^5$ at ±1.0 V, and a Schottky barrier height of 0.83 eV. A strong linear relationship between light intensity and photocurrent is also observed. Furthermore, the device exhibits a peak external quantum efficiency of ~26% at 540 nm and a peak internal quantum efficiency of ~97% at 410 nm. Transient photocurrent and photovoltaic measurements show approximately one-microsecond extraction and several-millisecond recombination times, respectively, revealing effective charge collection for photovoltaic applications. These results indicate that the rf-MLG/n-Si Schottky junction is well-formed and achieves performance comparable to that of SLG devices, demonstrating its potential for optoelectronic applications.

[*]Corresponding author. Email: hojun@hirosaki-u.ac.jp (H. J. Im)

# 1. Introduction

Photovoltaic devices based on graphene-silicon Schottky junctions have significantly advanced applications such as solar cells and photodetectors. These breakthroughs address long-standing issues associated with metal-semiconductor Schottky junctions, including high reflectivity, low transmittance of the metal layer, and high recombination rates at interfacial states.

In 2011, Chen et al. reported on Gr/n-Si Schottky photodetectors, demonstrating high responsivity [1]. Notably, the infrared sensitivity of these photodetectors surpasses that of their silicon-based counterparts [2–12]. These advantages arise from exceptional optical and electrical properties of graphene, including its Dirac cone band structure with a zero bandgap and the robust van der Waals heterostructure formed between graphene and silicon [13–22]. The Gr/n-Si Schottky solar cell was first demonstrated by X. Li et al. in 2010, achieving a power conversion efficiency (PCE) of 1.65% [13]. Since then, rapid advancements have been made, leading to PCE values as high as 17% in 2023 [23]. These improvements have been achieved through various strategies, including graphene hole doping, antireflection coatings, interfacial oxide layers, optimization of graphene layer number, and other approaches [14–17,23–32]. Furthermore, graphene-based devices generally exhibit superior performance compared to other 2D material-based devices, such as those using $MoS_2$ or $WSe_2$, which form heterojunctions with silicon and typically demonstrate PCEs of approximately 3-5% [33–36]. The advantage of graphene can be attributed to the high carrier mobility, easily tunable work functions, and a well-defined Schottky barrier with silicon. In contrast, semiconducting 2D materials typically form more complex p–n heterojunctions and suffer from lattice defects at the interface, leading to less favorable band alignment and higher interfacial recombination, respectively [36,37].

Despite these significant advancements, further efforts are needed to facilitate practical applications of Gr/n-Si Schottky junction devices. Key challenges include the handling of single-layer graphene (SLG) during fabrication due to its atomic-scale thickness and the presence of polymer residue on graphene such as polymethyl methacrylate (PMMA) which is formed during the transfer process and can adversely affect conversion efficiency [38,39]. PMMA residues are typically removed using an acetone solution and thermal annealing at approximately 400°C. However, this method restricts the use of polymer substrates, which are critical for flexible devices but susceptible to thermal degradation.

To address these issues, several research groups have explored polymer-free graphene transfer methods [40–43]. Additionally, challenges such as low conductivity, limited durability, and high reflectivity continue to hinder progress in SLG applications. Multilayer graphene (MLG) has emerged as a potential solution to these issues, offering improved robustness and conductivity due to its thicker layers [22,31]. However, increasing the number of stacked layers can lead to graphitization, and Bernal stacking, a common stacking arrangement, can degrade carrier mobility and two-dimensionality [18]. On the other hand, rotationally faulted stacking—also referred to as twisted, misoriented, or turbostratic stacking—weakens the bonding between graphene layers along the c-axis, thereby enhancing two-dimensionality and preserving electrical and optical properties similar to those of SLG, showing the so-called Moiré pattern [19,20,44–46]. This preservation occurs because the rotational mismatch electronically decouples the individual graphene sheets, maintaining the unique linear Dirac cone band structure and high carrier mobility characteristic of an isolated layer.

However, there have been very few reports on the photovoltaic effects in rotationally faulted multilayer graphene (rf-MLG)/n-Si Schottky optoelectronic devices so far. In this study, we demonstrate the fabrication of a rf-MLG/n-Si Schottky junction and evaluate its potential as a photovoltaic device. The rf-MLG layer, with a thickness of approximately 22 nm and a transmittance of 30% at 550 nm, was synthesized via chemical vapor deposition (CVD) on a 5 μm Ni foil catalyst and transferred to silicon using a PMMA-free method. We assessed the photovoltaic performance through J-V characteristics, quantum efficiency measurements, and transient photocurrent (TPC) and photovoltage (TPV) responses. Our results show that rf-MLG/n-Si Schottky junctions perform comparably to SLG devices in the J-V characteristics, EQE, and the transient responses.

## 2. Experimental details

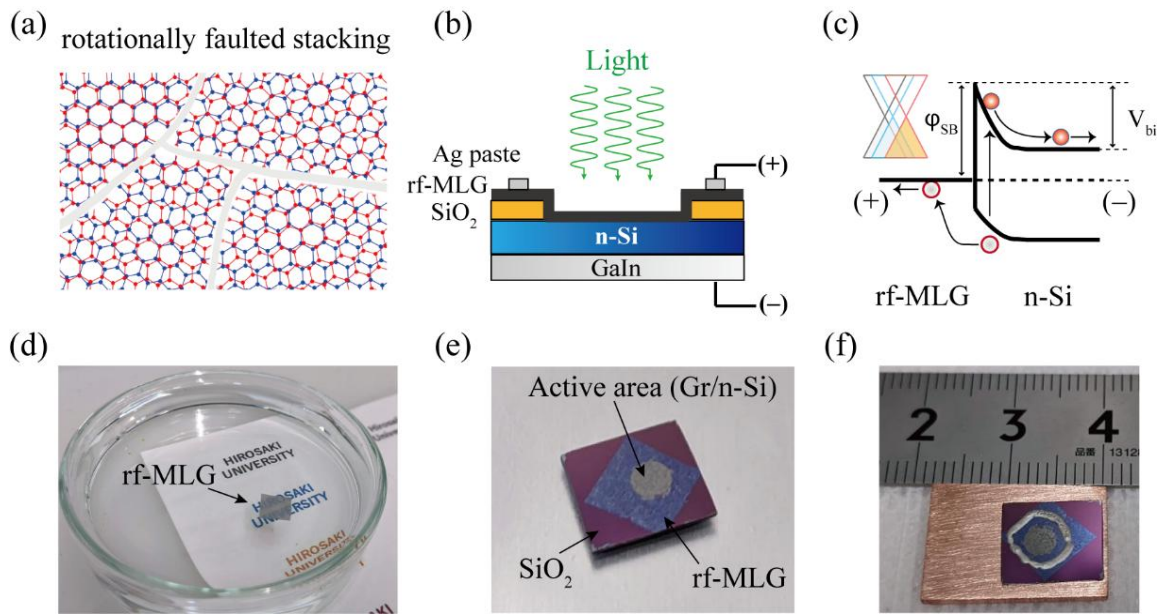

**Fig. 1.** (a) Schematic of the rotationally faulted two graphene layers with several domains, where various Moiré patterns appear due to different rotational angles. (b) Cross-sectional view of rf-MLG/Si Schottky junction device. rf-MLG plays a role as metal and transparent electrode layers at the same time. (c) Energy diagram of rf-MLG/Si device for the short circuit. $\phi_{SB}$ and $V_{bi}$ represent Schottky barrier potential and built-in potential, respectively. (d) Photograph of a free-standing rf-MLG without the PMMA support layer. (e, f) Fabricated rf-MLG/n-Si Schottky junction device.

rf-MLG was synthesized on a Ni foil catalyst using low-pressure CVD, as described in our previous report, where the thickness of the MLG was controlled by the thickness of the Ni foil [22]. The Ni foil was first annealed at 1000 °C for 1 hour in a vacuum of ~$10^{-6}$ Torr. A gas mixture of $CH_4$ (50 sccm), $H_2$ (10 sccm), and Ar (100 sccm) was then introduced into the CVD quartz chamber for 7 minutes at ~$10^{-1}$ Torr. Finally, the furnace was fully opened to quench-cool the sample to room temperature in an argon atmosphere. For this study, we used a 5 μm thick Ni foil to produce thin MLG, which can be transferred on a target substrate without a PMMA support layer. This prevents degradation from PMMA residues and enables cost-effective fabrication. This method is also suitable for polymer substrates sensitive to heat, as it does not need thermal annealing at around 400°C to remove the PMMA residues. To fabricate the rf-MLG/n-Si Schottky junction device, we employed an n-Si (100) substrate with a resistivity of approximately 10 Ωcm, coated with a thermal oxide ($SiO_2$) layer of about 300 nm on both sides. The thermal oxide layers were removed using a buffered oxide etchant for 7 minutes. A circular window with a 3.4 mm diameter was defined as the active area by removing the thermal oxide layers using a buffered oxide etchant solution and a polyvinyl tape mask. During the transfer process, the rf-MLG floated without the need for a PMMA support layer due to the robustness of the thicker graphene layers, allowing for a PMMA-free transfer onto the n-Si substrate, as shown in Fig. 1(d). This step simplifies fabrication process compared to that of the SLG/n-Si Schottky device. The rf-MLG/n-Si device was then completed by forming electrodes on both the front and back sides. The back ohmic contact was formed using GaIn eutectics on the n-Si, while the front contact was made with Ag paste applied to the graphene. Figures 1(b) and 1(f) show a schematic diagram and a photograph of the fabricated rf-MLG/n-Si Schottky junction device, respectively.

The surface morphology of rf-MLG was examined using scanning electron microscopy (SEM) with a JSM-7000F instrument (JEOL Ltd.) as shown in Fig. S1. To characterize the rf-MLG transferred onto the $SiO_2$/n-Si substrate [47], a micro-Raman spectrometer (NRS-5100,

Japan Optics Ltd.) with a 532 nm excitation laser was employed. Current density-voltage (J-V) characterization was performed using a Keithley 2400 source meter (Keithley Instruments Inc.) under both dark and illuminated conditions, with light intensity calibrated to 1 sun (AM1.5G) using a Si-detector. External quantum efficiency (EQE) spectra were recorded using a monochromator (CS260, Oriel Ltd.), a lock-in amplifier (LI5640, NF), and a Xenon lamp as the light source, with the monochromatic light size being approximately 1 mm$^2$. Transmittance measurements were conducted using a monochromator and an integrating sphere detector. Atomic Force Microscopy (AFM) measurements were carried out using a NanoNavi2/E-Sweep (SII Nano Technology Inc.) to evaluate the thickness of rf-MLG. The TPC and TPV responses were obtained using a 639 nm laser diode with the output power of 10 mW (HL6358MG, Thorlabs) and an oscilloscope (MSO5354, Rigol).

## 3. Results and discussion

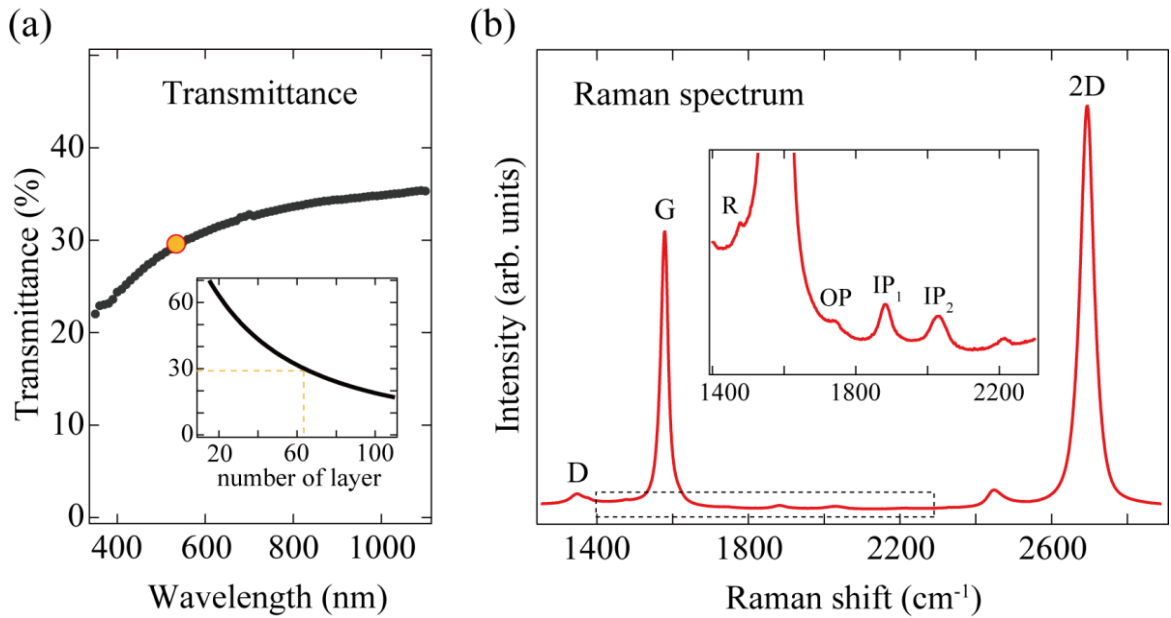

**Fig. 2.** (a) Transmittance of rf-MLG in the region of 350–1100 nm. Inset shows the relationship between the number of graphene layers and transmittance at 550 nm. (b) Raman spectrum of rf-MLG transferred onto the SiO$_2$/n-Si substrate. Inset shows the enlarged Raman spectrum in the range of the dashed square.

We investigated the crystallinity of the rf-MLG/n-Si Schottky junction device using spectroscopic techniques and evaluated its performance through J–V characteristics, EQE, and transient response measurements.

Figure 2(a) shows the transmittance of graphene transferred onto a glass substrate. Transmittance slightly increases from 22% to 35% as the wavelength changes from 350 nm to 1100 nm. The thickness of graphene can be determined based on the known relationship between the number of layers and the transmittance at 550 nm ($T = (1 + 1.13\pi\alpha N/2)^{-2}$, where N is the number of graphene layer, $\alpha$ is the fine-structure constant, and 1.13 is the correction coefficient of the universal optical conductance) [22,48]. The rf-MLG used in this study exhibits a transmittance of about 30% at 550 nm, corresponding to around 65 layers, as indicated in the inset of Fig. 2(a). Given the interlayer spacing of 0.34 nm, the thickness of the rf-MLG is estimated to be approximately 22 nm. This result demonstrates that a thin rf-MLG layer was successfully obtained using a 5 μm-thick Ni catalyst foil, in comparison with previous studies where thicknesses were approximately 44 nm and 212 nm when using 10 μm and 50 μm Ni foils, respectively [22]. The rf-MLG thickness was also evaluated by AFM, which revealed a step height of approximately 35 nm (Fig. S2 (b)). Although this value is slightly higher than the optical estimate, it remains consistent. The difference between the optically estimated average thickness (~22 nm) and the locally measured AFM thickness (~35 nm) may be attributed to surface inhomogeneity in the transferred rf-MLG layer. Figure S2(a) displays the variation in the surface height of rf-MLG on n-Si across different regions.

Raman spectroscopy is widely regarded as a powerful method for analyzing the crystalline quality and stacking characteristics of rf-MLG. The crystallinity is typically assessed by observing the G, 2D, and D bands. In high-quality SLG, the intensity ratio of the 2D to G bands ($I_{2D}/I_G$) is generally greater than 2, and the full width at half maximum (FWHM) of the 2D band is less than 40 cm$^{-1}$ [49]. Figure 2(b) shows a representative Raman spectrum of rf-MLG, transferred onto the SiO$_2$/n-Si substrate, exhibiting the G, 2D, and D bands at approximately 1579, 2694, and 1350 cm$^{-1}$, respectively. The spectrum of rf-MLG reveals the $I_{2D}/I_G$ ratio of around 1.5, indicating SLG-like behavior due to the rotational stacking of graphene layers, despite the large number of layers in rf-MLG. Additionally, the 2D band displays a single symmetric peak, indicating rotational stacking rather than non-rotational or Bernal stacking, where the 2D peak is broader and asymmetrical [50]. It has been reported that rotational misalignment greater than 7 degrees is known to produce a single Lorentzian peak [19,20,51]. The D band, which arises from defects in the crystal structure, provides insight into the defect density through the intensity ratio of the D to G bands ($I_D/I_G$) [22,49,52]. The small $I_D/I_G$ ratio of approximately 0.04 and the FWHM of the 2D band (~50 cm$^{-1}$) suggest that rf-MLG exhibits high crystalline quality. Further evidence of rotational stacking in rf-MLG is seen in the weak Raman peaks between ~1400 and ~2300 cm$^{-1}$, as shown in the inset of Figure 2(b). The in-plane (IP) and out-of-plane (OP) modes refer to vibrations of carbon atoms within the graphene plane and perpendicular to it, respectively. The strong IP modes at approximately 1882 and 2030 cm$^{-1}$, compared to the weaker OP mode around 1745 cm$^{-1}$, highlight the strong two-dimensionality of rf-MLG. It is important to note that the OP mode at 1745 cm$^{-1}$ is typically an infrared-active mode in Bernal stacking and is rarely observed in Raman spectra. Thus, the presence of the OP mode suggests that the stacking of rf-MLG differs from Bernal stacking. Additionally, the so-called R peak near the G band (~1477 cm$^{-1}$), associated with superlattice scattering caused by rotational stacking, are observed [53–58]. It should also be mentioned that the rf-MLG used in this study is polycrystalline, and its Raman spectra exhibit both SLG- and few-layer graphene-like characteristics depending on the position [22].

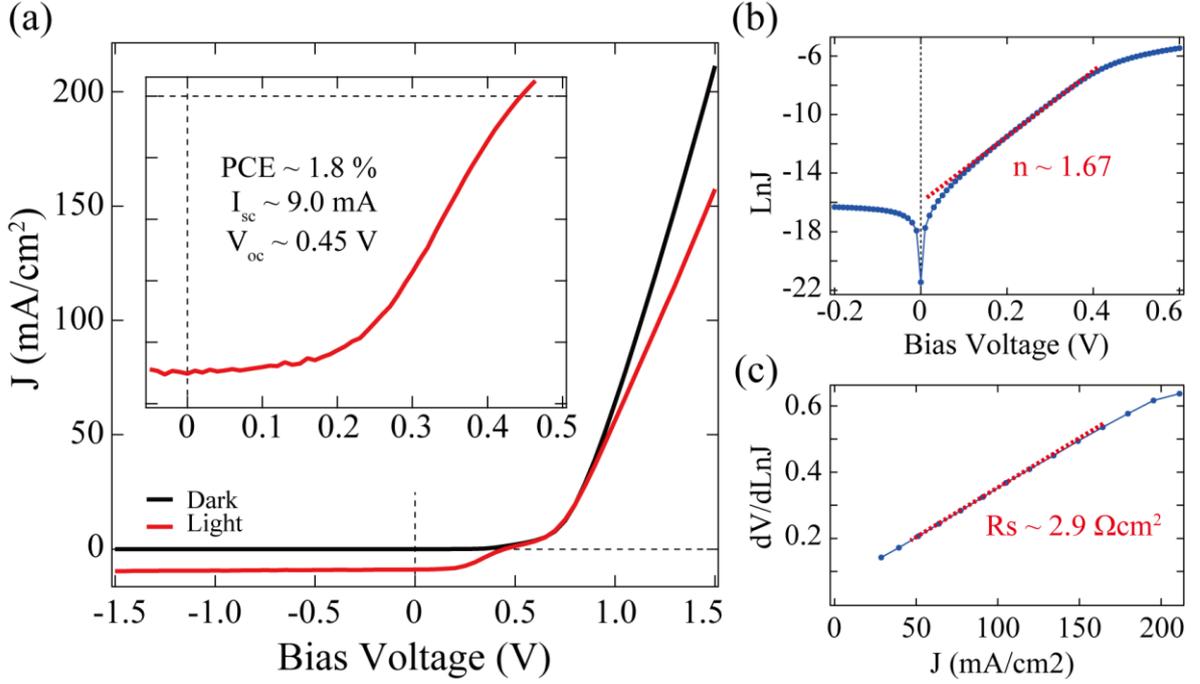

**Fig. 3.** (a) J-V curve of the rf-MLG/n-Si Schottky junction device with (Light) and without (Dark) the illumination. The inset shows an enlarged J-V curve in the Light. (b) J-V curve in the dark is plotted in a semi-logarithmic scale. (c) The series resistance is obtained from the dV/dLnJ versus J plot. The dashed lines are guides to the eye.

Figure 3 (a) shows the J-V characteristics under the illumination of 1 SUN and in the dark. The dark J-V curve (black line) reveals a good rectifying behavior with the rectification ratio of ~4.3 × 10$^5$ at ±1.0 V, indicating a good formation of the Schottky junction between the rf-MLG and n-Si surfaces. In the Schottky junction, J-V characteristic can be expressed by the diode equation based on the thermionic-emission as in equations (1) and (2):

$$J = J_0 \left( e^{qV/nkT} - 1 \right) \quad (1)$$

$$J_0 = A^* T^2 e^{-q\phi_{SB}/kT} \quad (2)$$

where J is diode current density, $J_0$ reverse saturation current, q electronic charge, V applied voltage across the diode, $\phi_{SB}$ Schottky barrier potential, $A^*$ effective Richardson constant [59]. Here, we used 112 A/cm$^2$K$^2$ as the $A^*$ value of n-Si following the previous reports [13,38,39].

From the above equations, φ$_{SB}$ and ideality factor (n) are given by eq. (3) and (4), respectively.

$$\phi_{SB} = \frac{kT}{q} ln\left(\frac{A^*T^2}{J_0}\right) \quad (3)$$

$$n = \frac{q}{kT}\frac{dV}{dlnJ} \quad (4)$$

The value of J$_0$ was estimated to be ~9.5 × 10$^{-8}$ A by extrapolating the fitted line (red dashed line) to zero voltage in Fig. 3(b). From eq. (4), the value of n can be determined from the slope of the J-V characteristic in semi-logarithmic scale and found to be approximately 1.67 in the bias region of 0.2–0.3 V as shown in Fig. 3(b). This is comparable to that of SLG/n-Si Schottky junction [38,39] and is also consistent with that of a flake-HOPG/Si Schottky junction diode where the outermost layer of the graphite has been considered to play a role as single graphene sheet [60]. The value of φ$_{SB}$ was estimated to be ~0.83 V. Here, it should be mentioned that the slope in the high bias voltage region (0.5–0.7 V) is different with that of the low bias voltage region (0.1–0.3 V), indicating a different process of the Schottky diode behavior in the low and high bias region. Series resistance was obtained from the slope of the dV/dLnJ vs. J plot in forward-bias region and was estimated to be ~ 2.9 Ωcm$^2$ (Fig. 3(c)) [61]. These values are also comparable to the performances of the Schottky junction between the SLG and n-Si substrate: in the previous reports, n, φ$_{SB}$, and Rs were estimated to 1.3, 0.82 V, and be ~ 4.7 Ωcm$^2$ [38,39].

Fig. 3(a) and its inset show the J-V characteristic under the illumination of 1 sun (red line). We observe that the photocurrent flows in the opposite direction to the bias potential. The open circuit voltage (V$_{OC}$), short circuit current density (J$_{SC}$), and fill factor (FF) are estimated to be 0.45 V, 9.0 mA/cm$^2$, and 46%, respectively. And these values result in the PCE of 1.8 %, which is comparable to the first Gr/n-Si solar cell of 1.65% [13], and superior to the first diode of SLG/n-Si [1]. Recently, a state-of-art pristine SLG/n-Si solar cells show about 5 % with J$_{SC}$ ~ 30 mA/cm$^2$ and V$_{OC}$ ~ 0.45 V [23]. When we consider the low transmittance of rf-MLG of ~30 % at 550 nm, which is around one third of the SLG transmittance (~97 %) [62], the performance of rf-MLG is considered reasonable. This implies that the Schottky junction of the rf-MLG/n-Si device well plays a role as good as that of the SLG/n-Si device. For comparison, the photovoltaic parameters of the representative pristine Gr/n-Si solar cells are summarized in table 1. On the other hand, we can observe a prominent S-shape feature near $V_{oc}$. This feature influences the J-V curve in the 0.25-0.6 V range, resulting in a reduction in both the FF and PCE. It has been considered to arise from interfacial barriers [16] and/or a trap of photocurrent [38], both of which lead to recombination losses. It has been reported that the optimization of insulating interlayer thickness [6], the hole-doping of graphene [14] and the removal of PMMA [7] can eliminate the S-shape feature, improving FF and PCE.

**Table 1**

Comparison of the photovoltaic parameters of this work with those of representative pristine Gr/n-Si Schottky junction solar cells.

| Graphene | PCE (%) | $V_{OC}$ (V) | $J_{SC}$ (mA/cm$^2$) | FF (%) | n | $\phi_{SB}$ (eV) | Ref. |
|---|---|---|---|---|---|---|---|
| rf-MLG (~65 layers) | 1.8 | 0.45 | 9 | 46 | 1.67 | 0.83 | [a] |
| MLG (a few layers) | 1.65 | 0.48 | 6.5 | 56 | 1.57 | 0.78 | [13][b] |
| SLG | 1.9 | 0.42 | 14.2 | 32 | 1.6 | 0.79 | [14] |
| SLG | 2.66 | 0.43 | 16.15 | 38.52 | 3.79 | N/A | [21] |
| SLG | 3.9 | 0.41 | 29.2 | 33 | 1.3 | 0.82 | [38] |
| SLG | 2.8 | 0.39 | 25.4 | 28 | 1.16 | 0.78 | [39] |
| SLG | 4 | 0.4 | 26.9 | 37 | N/A | N/A | [23] |
| MLG (3 layers) | 7.3 | 0.415 | 38.8 | 45 | N/A | N/A | [23][c] |

Footnotes: [a] This work. [b] First report. [c] Current record.

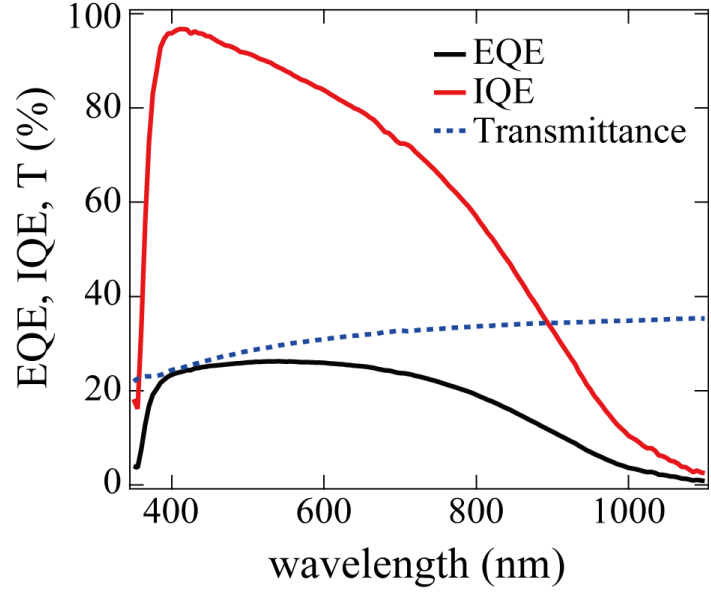

**Fig. 4.** EQE spectrum of the rf-MLG/Si Schottky junction device in the range of 350−1100 nm. IQE spectrum is obtained by dividing the EQE spectrum by the transmittance.

Figure 4 shows the EQE and internal quantum efficiency (IQE) spectra in the range of 350 to 1100 nm. The EQE exhibits low values across the entire range, with a maximum of approximately 26% at around 540 nm, due to the low transmittance. Notably, the peak at 540 nm is close to wavelength corresponding to the maximum intensity of the solar spectrum (approximately 500 nm). In the short-wavelength region (less than 400 nm), the EQE decreases significantly, a trend typically attributed to the surface recombination. Between 400 and 650 nm, the EQE spectrum remains relatively flat. In the long-wavelength region (650 to 1100 nm), the EQE decreases as the wavelength increases. This decline is governed by the absorption edge of n-Si, which corresponds to its 1.1 eV band gap. These overall EQE behaviors are similar to those of the SLG/n-Si solar cell [38].

We can estimate the integrated $J_{SC}$ by using the following equation:

$$J_{sc} = q \int_{350nm}^{1100nm} EQE(\lambda) \cdot \phi(\lambda) d\lambda \qquad (5)$$

where $\phi(\lambda)$ is the photon flux of AM1.5G [63]. The integrated $J_{SC}$ is about 7.9 mA/cm² which is almost consistent with the value of $J_{SC}$ (9.0 mA/cm$^2$) obtained from J-V measurements.

To evaluate the photogeneration originating from the Schottky junction, we need to extract IQE, which excludes absorption in the rf-MLG layer and only accounts for the photon arriving at the interface between rf-MLG and n-Si. To this end, we here define IQE as the EQE divided by the transmittance. It is impressive that the IQE near 410 nm is close to 100% (approximately 97%) and is approximately 89% at 540 nm. This observation indicates that the Schottky junction between rf-MLG and n-Si efficiently generates photocurrents, particularly in the violet-to-green spectral region (380-570 nm).

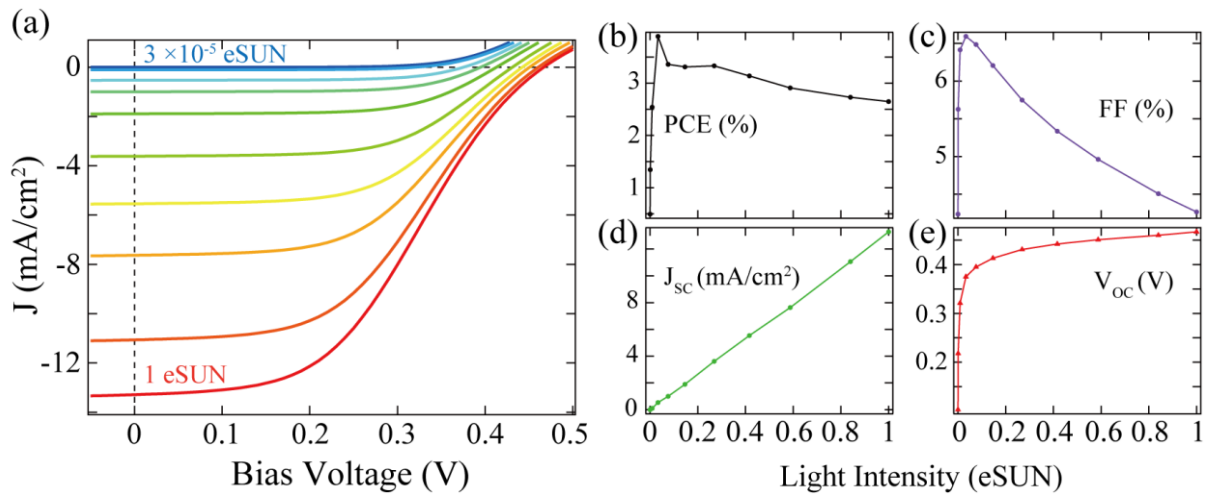

**Fig. 5.** (a) Light intensity dependency of the J-V characteristics of rf-MLG/n-Si Schottky junction device. (b-e) Plots of PCE, FF, $J_{SC}$, and $V_{OC}$ as a function of the light intensity.

Figure 5(a) shows the J-V characteristics under the illumination of the halogen lamp from an equivalent $3 \times 10^{-5}$ to 1 SUN. The equivalent SUN (eSUN) was estimated from the photocurrent of a calibrated Si detector: We here define the 1 eSUN as a light intensity of the halogen lamp which has the same photocurrent of the calibrated Si detector under a 1 SUN light intensity of a Xenon lamp. We recognize that the $J_{SC}$ of halogen lamp is larger than that of Xenon lamp (Fig. 3(a)). This may come from the difference of spectral weight as a function of the wavelength between the halogen and the Xenon lamps. The detailed origin should be studied more. In this study, we focus on the variation of photovoltaic parameters as a function of the light intensity. Figure 5(d) shows that the $J_{SC}$ values are linearly proportional to the light intensity from $3 \times 10^{-5}$ to 1 eSUN. The $V_{OC}$ suddenly increases around 0.01 eSUN and eventually closes to 0.47 V around the 1 eSUN as the light intensity increases (Fig. 5(e)). Such behaviors indicate that the rf-MLG/n-Si Schottky junction device functions effectively as a photodiode. It is likely that the PCE and FF have correlated to each other as shown in Fig. 5(b) and Fig. 5(c): Above 0.03 eSUN, they gradually decrease as the light intensity increases. On the other hand, below 0.03 eSUN, they abruptly decrease with the light intensity. These indicate that the device has the best performance in weak light intensity (~0.03 eSUN).

Figure 6 shows the TPV and TPC responses of the rf-MLG/n-Si Schottky device under a 639 nm square pulse at 15 Hz. The laser with the output power of 10 mW was used and its intensities were controlled by using neutral density filters with the optical density (OD) 1, 2, and 3 which correspond to intensities of 0.1, 0.01, and 0.001 times the original intensity (OD0) without a filter. The decay times of the TPC and TPV response can be interpreted as the charge extraction time and the recombination time of photoexcited electron-hole pairs, respectively. The relaxation times for TPC and TPV were estimated to be approximately one microsecond and several milliseconds, respectively. The faster decay of the TPC compared to the TPV suggests that the photogenerated current is effectively collected by the electrodes in a photovoltaic device. These behaviors were observed even under weak laser pulse intensity of OD3. Additionally, the overall responses of both TPC and TPV show little change on the falling edge, although the TPV response slows slightly on the rising edge (Fig. S3). For a more detailed analysis, the TPV responses were fitted using a double exponential decay equation. The two relaxation times were estimated to be $\tau_1 = 2.8 \times 10^{-4}$ s and $\tau_2 \sim 4.4 \times 10^{-3}$ s without the neutral density filter (OD0). These values are in good agreement with previous reports, where $\tau_1$ and $\tau_2$ were attributed to the interface of the Schottky junction and the Si bulk, respectively. As the laser intensity decreases from OD0 to OD3, $\tau_1$ doubles, reaching up to $6.0 \times 10^{-4}$ s, while $\tau_2$ shows little change (Table S1). This indicates that carrier recombination at the interface slows down as the light intensity decreases, whereas the bulk recombination process is largely unaffected by light intensity. This is further supported by the J-V characteristics, where both FF and PCE improve as the light intensity decreases above 0.03 eSUN, as shown in Fig. 5. Compared to conventional SLG/n-Si solar cells, where PMMA is used as a support layer during the transfer process and $\tau_2$ reflects contributions from both the Si bulk and the capacitive effects of PMMA residues [38], we employed a PMMA-free transfer process in this study. This allows us to attribute the observed TPV relaxation time solely to the bulk properties of n-Si. In the TPC responses, the rf-MLG/n-Si Schottky junction device does not exhibit the issues caused by PMMA residues in SLG/n-Si solar cells, such as the rapid decay of photocurrent at the rising edge due to parasitic current leakage and the shoulder at the falling edge caused by photocurrent trapping [38].

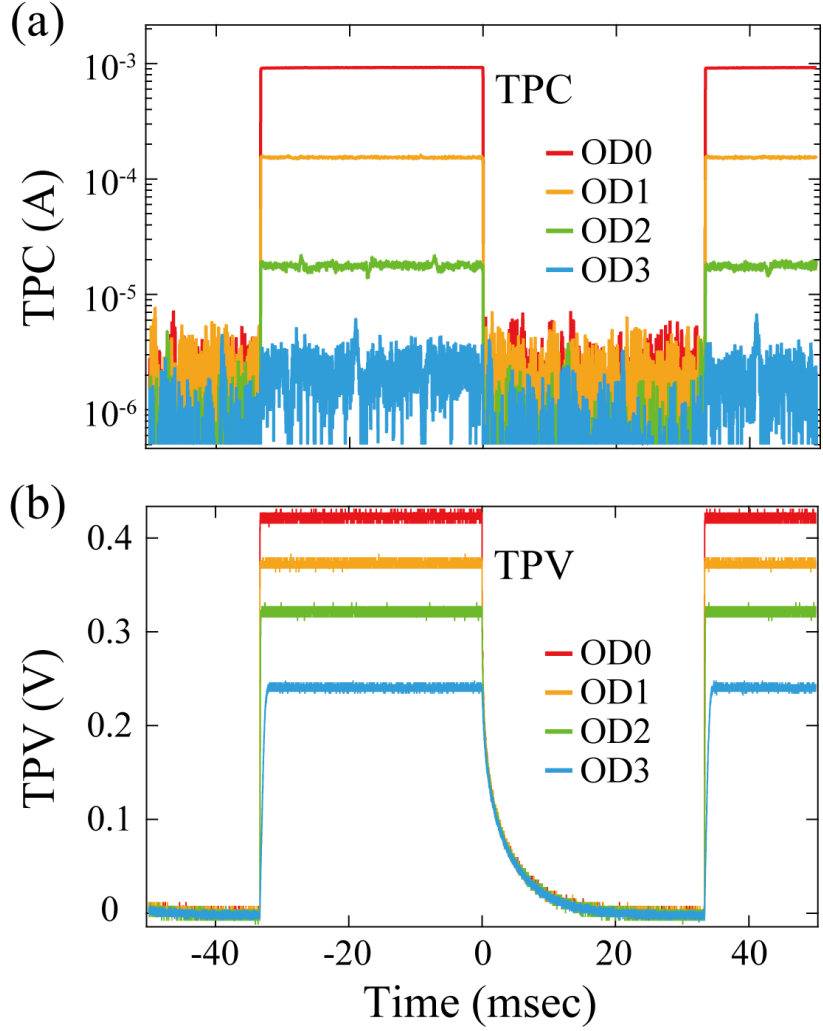

**Fig. 6.** (a) TPC and (b) TPV responses of the rf-MLG/n-Si Schottky junction device under a 639 nm square laser pulse. For clarity, the time at the falling edges is set to 0 s. Laser intensities were controlled using neutral density filters with optical densities of OD1, OD2, and OD3, corresponding to 0.1, 0.01, and 0.001 times the original intensity (OD0) without a filter.

## 4. Conclusion

We report on the fabrication and photovoltaic performance of an rf-MLG/n-Si Schottky junction device. The rf-MLG with an optically estimated average thickness of approximately 22 nm was synthesized using a 5 μm Ni foil catalyst via CVD. To prevent degradation from PMMA residues and simplify fabrication, a PMMA-free transfer process was used to transfer the graphene onto the silicon substrate. The device exhibited excellent J-V characteristics, with the ideality factor of 1.67, the rectification factor of ~$4.3 \times 10^5$ at ±1.0 V and the Schottky barrier height of 0.83 eV. These are comparable to those of SLG devices. A linear relationship between light intensity and photogenerated current confirmed the potential as an efficient photodetector. EQE measurements revealed a peak value of 26% at 540 nm, with a broad response range from 400 to 900 nm. Notably, the value of IQE approached 97% near 410 nm. TPC and TPV measurements showed a short carrier extraction time of approximately one microsecond and a long carrier recombination time in the millisecond range, respectively, both of which are desirable for solar cell and photodetector applications. These results confirm that rf-MLG/n-Si Schottky junctions possess outstanding diode characteristics, underscoring their promise as a robust platform for photovoltaic technology.

# Supplementary materials

**Photovoltaic Performance of a Rotationally Faulted Multilayer Graphene/n-Si Schottky Junction**

Hojun Im[*] and Masahiro Teraoka

*Graduate School of Science and Technology, Hirosaki University, Hirosaki 036-8561, Japan*

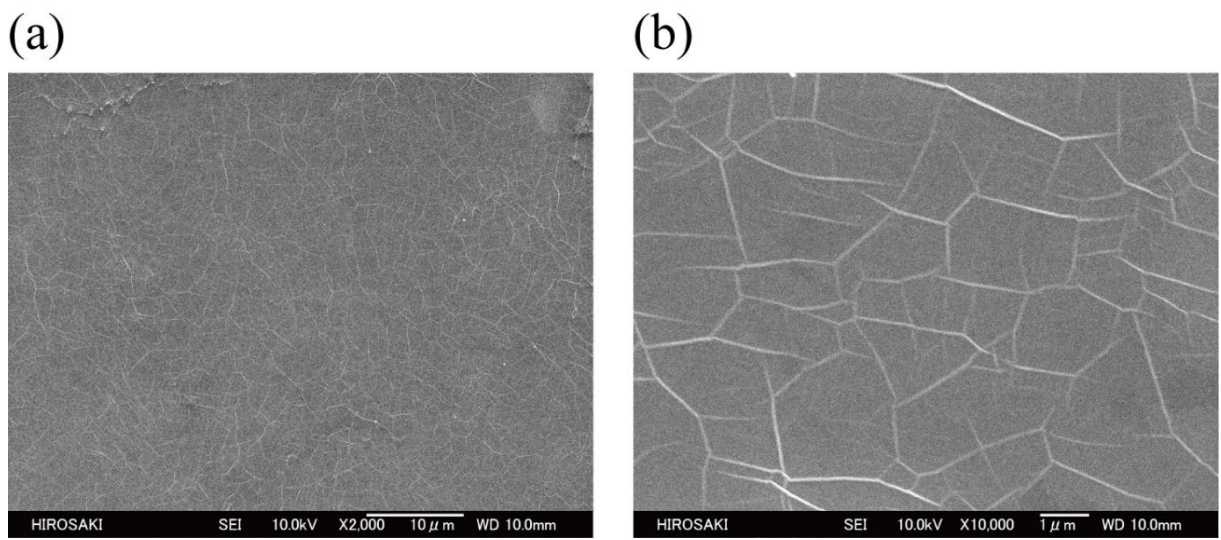

**Fig. S1.** SEM images of rf-MLG transferred onto a bare n-Si substrate at (a) ×2000 and (b) ×10000 magnifications. The white wrinkles, formed by the shrinkage of the MLG sheets during the cooling in the CVD process, reveal the layered structure of the material.

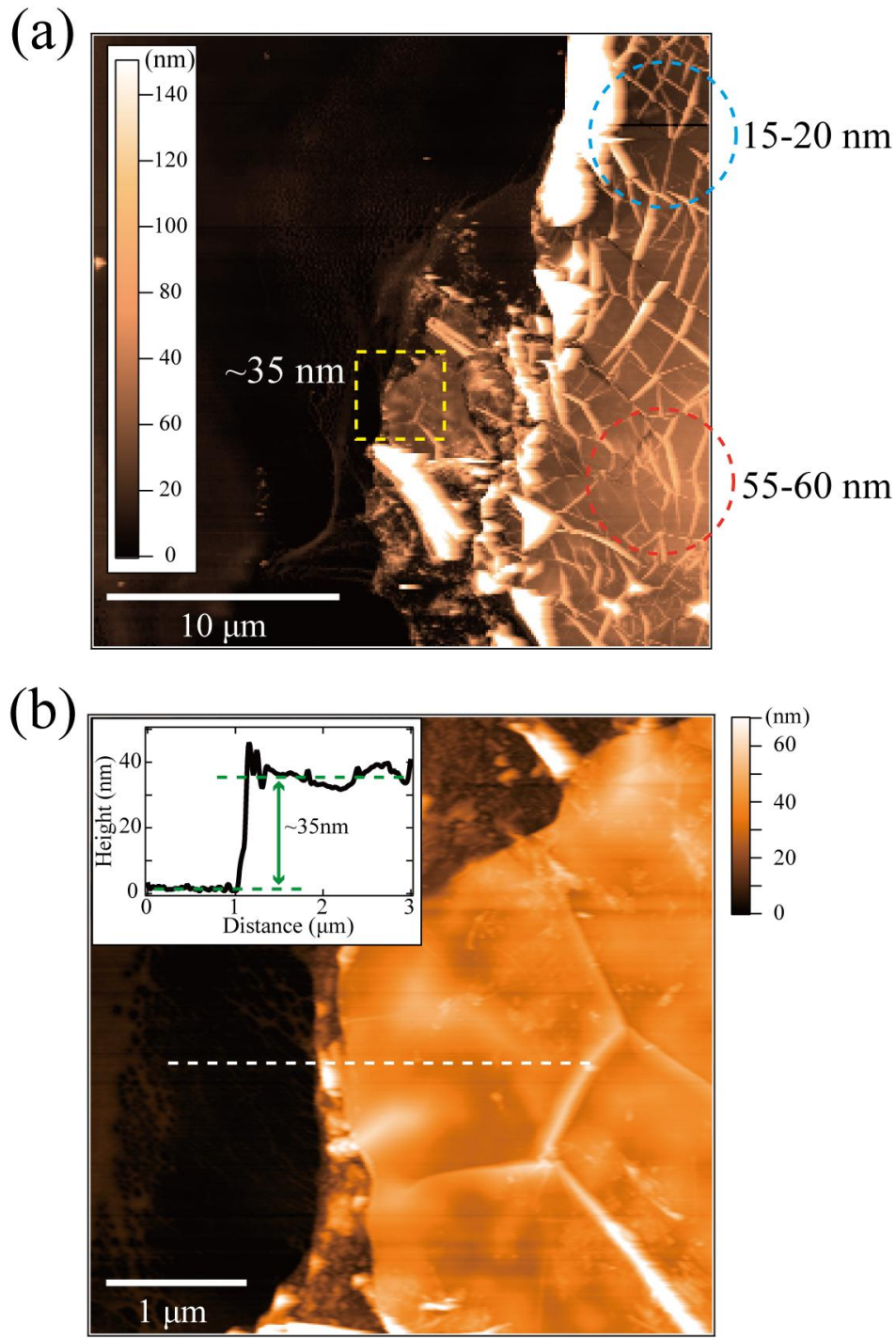

**Fig. S2.** AFM images of rf-MLG, transferred onto a bare n-Si substrate, at the edge. (a) Large-area AFM image (scale bar = 10 μm); the dashed square and circles show regions of inhomogeneous surface height. (b) High-resolution AFM image of the area marked by the dashed square in (a) (scale bar = 1 μm). The corresponding height profile along the dashed line, shown in the inset, indicates a step height of approximately 35 nm.

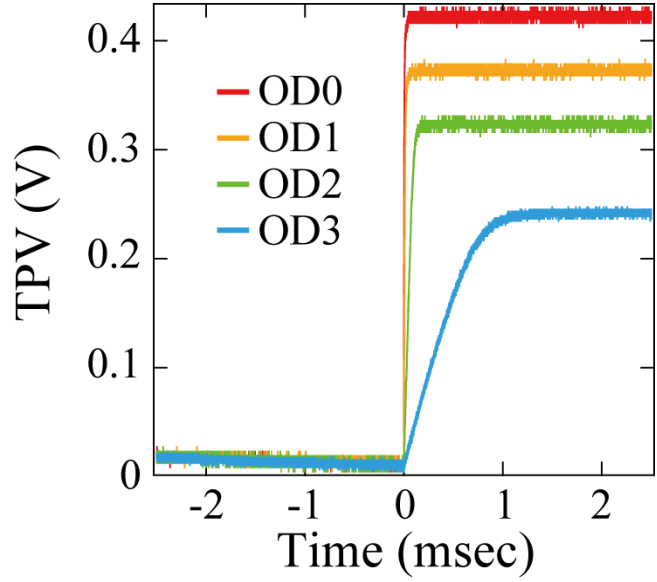

**Fig. S3.** Enlarged plot of the laser intensity-dependent TPV responses on the rising edge of the rf-MLG/n-Si Schottky junction device. For clarity, the time at the rising edges was set to 0 s.

**Table S1**

Extracted relaxation times of TPV responses under various laser intensities, controlled by neutral density filters with different optical densities (OD).

| Light intensity (Optical density) | $\tau_1$ ($\times 10^{-4}$ sec) | $\tau_2$ ($\times 10^{-3}$ sec) |
|---|---|---|
| 1 (OD0) | 2.8 | 4.4 |
| 1/10 (OD1) | 3.1 | 4.3 |
| 1/100 (OD2) | 4.0 | 4.6 |
| 1/1000 (OD3) | 6.0 | 4.8 |